\documentclass[english,journal=jctcce,manuscript=article,email=true,etalmode=truncate,maxauthors=0,doi=true]{achemso}

\usepackage{soul}
\usepackage{graphicx}
\graphicspath{{figures/}}
\usepackage{physics}
\usepackage{multirow}
\usepackage{xcolor}
\usepackage{caption}
\usepackage{subcaption}
\usepackage{bm}
\usepackage{amsfonts}
\usepackage{yfonts} % for \textgoth
\usepackage[cal=boondoxo,bb=stix]{mathalpha}  % for \mathcal{lowercase letters} and \mathbb{lowercase letters}
\usepackage[title,titletoc]{appendix}

\usepackage{diagbox}
\usepackage{multicol}
\usepackage{booktabs} % for tables obtained from pandas to_latex function
\usepackage{threeparttable}
\usepackage{float}

\usepackage{xspace} % for \xspace in macros

\usepackage{longtable}

\usepackage{siunitx}
\usepackage{hyperref} % hyperref must be placed before cleveref
\hypersetup{colorlinks=true, citecolor=blue, urlcolor=blue, linkcolor=blue}
\usepackage{cleveref}
  	\crefname{figure}{Figure}{Figures}
  	\crefname{table}{Table}{Tables}
  	\crefname{equation}{Eq.}{Eqs.}
  	\crefname{section}{Section}{Sections}
  	\crefname{subsection}{Section}{Sections}
  	\crefname{subsubsection}{Section}{Sections}
  	\crefname{algorithm}{Algorithm}{Algorithms}
    \crefname{appendix}{Appendix}{Appendices}
\newcommand{\doi}[1]{\href{http://dx.doi.org/#1}{\nolinkurl{#1}}}

\newcommand{\code}[1]{\texttt{#1}}
\newcommand\vartextvisiblespace[1][.5em]{%
  \makebox[#1]{%
    \kern.07em
    \vrule height.3ex
    \hrulefill
    \vrule height.3ex
    \kern.07em
  }% <-- don't forget this one!
}

\usepackage{todonotes}

\title{SAP-X2C: Optimally-Simple Two-Component Relativistic Hamiltonian With Size-Intensive Picture Change} %Title of paper

\author{Kshitijkumar A. Surjuse}
\author{Edward F. Valeev}
\affiliation{Department of Chemistry, Virginia Tech, Blacksburg, VA 24061}
\email{efv@vt.edu}

\begin{document}

\date{\today}
\newpage

\begin{abstract}
We present a simple relativistic exact 2-component (X2C) Hamiltonian that models  two-electron picture-change effects using Lehtola's superposition of atomic potentials (SAP) [S. Lehtola, {\em J. Chem. Theory Comput.} {\bf 15}, 1593–1604 (2019)]. The SAP-X2C approach keeps the low-cost and technical simplicity of the popular 1-electron X2C (1eX2C) predecessor, but is significantly more accurate and has a well-defined thermodynamic limit, making it applicable to extended systems (such as large molecules and periodic crystals). The assessment of the SAP-X2C-based Hartree-Fock total and spinor energies, spin-orbit splittings, equilibrium bond distances, and harmonic vibrational frequencies suggests that SAP-X2C is similar to the more complex atomic mean-field (AMF) X2C counterparts in its ability to approximate the 4-component Dirac-Hartree-Fock reference. 
\end{abstract}

\maketitle %\maketitle must follow title, authors, abstract

\section{Introduction}\label{sec:intro}

Relativistic quantum chemistry\cite{KAS:grant:1988:AiAaMP,KAS:dyall:2007:,KAS:reiher:2009:}
is an approximation of quantum electrodynamics\cite{VRG:feynman:1949:PR}
that combines Dirac 1-particle kinematics with the classical description of electromagnetic interactions in a many-particle system.
By accounting for the dominant relativistic effects that are relevant for the chemistry and low-energy physics of atomistic matter and preserving the structure of nonrelativistic quantum many-body theory, accurate practical applications have become possible\cite{KAS:grant:1988:AiAaMP,KAS:eliav:1994:IJQC,KAS:visscher:1994:CPC,KAS:saue:1997:MP,KAS:yanai:2001:JCP,KAS:liu:2003:JTCC,KAS:dyall:2007:,KAS:saue:2011:C,KAS:kelley:2013:JCP,KAS:saue:2020:JCP,KAS:zhang:2020:JCP,KAS:repisky:2020:JCP,KAS:williams-young:2020:WCMS}.
Despite the close resemblance to the nonrelativistic quantum chemistry, the  appearance of positron-like degrees of freedom in the theory, such as bispinors (instead of spinors) and negative energy states, complicates both the formal and practical aspects of relativistic quantum chemistry.
Thus, from the earliest days, there was keen interest in constructing electron-only (2-component, 2C) heuristics that accurately approximate the reference 4-component (4C) treatment.
So-called quasi-relativistic 2C methods\cite{KAS:vanlenthe:1996:IJQC,KAS:reiher:2006:TCA,KAS:liu:2010:MP,KAS:saue:2011:C,KAS:barysz:2001:JMST,KAS:kutzelnigg:2012:CP,KAS:peng:2012:TCA,KAS:nakajima:2012:CR,KAS:knecht:2022:JCP,KAS:wang:2025:CPR} include: the Foldy-Wouthuysen (FW)\cite{KAS:foldy:1950:PR} transformation, the Douglas-Kroll-Hess (DKH)\cite{KAS:douglas:1974:AP,KAS:nakajima:2000:CPL,KAS:nakajima:2000:JCP,KAS:wolf:2002:JCP,KAS:vanwullen:2004:JCP,KAS:reiher:2004:JCP,KAS:reiher:2004:JCPa,KAS:wolf:2006:JCP,KAS:wolf:2006:JCPa,KAS:reiher:2007:PLA,KAS:peng:2009:JCP} methods, the zeroth-order regular approximation (ZORA)\cite{KAS:chang:1986:PS,KAS:lenthe:1993:JCP,KAS:lenthe:1993:JCP,KAS:vanlenthe:1994:JCP}, Barysz-Sadlej-Snijders (BSS)\cite{KAS:barysz:1997:IJQC,KAS:barysz:2001:JMST,KAS:barysz:2002:JCP} method, and exact-two-component (X2C) methods\cite{KAS:kutzelnigg:2005:JCP,KAS:liu:2009:JCP} are generalized forms of the one-step normalized elimination of small component (NESC) approach prescribed by Dyall\cite{KAS:dyall:1997:JCP,KAS:dyall:1998:JCP,KAS:dyall:1999:JCP,KAS:dyall:2001:JCP,KAS:dyall:2002:JCC} and self-consistent decoupling procedure\cite{KAS:liu:2007:JCP}. 
Our work adds to the X2C \emph{family} of methods\cite{KAS:kutzelnigg:2005:JCP,KAS:kutzelnigg:2006:MP,KAS:liu:2006:JCP,KAS:ilias:2007:JCP,KAS:liu:2009:JCP,KAS:konecny:2016:JCTC}, whose common trait is that the Fock-space Hamiltonian $H$ and other property operators expressed in a finite basis are rotated by a unitary matrix $U$ that is designed to block diagonalize the 1-particle Hamiltonian $h$ or its effective Fock-like equivalent (see \cref{sec:notation} for notation). Subsequently, the rotated Hamiltonian is projected onto the upper (electron-like) components of the bispinor:
\begin{align}
H_\text{X2C} = P_\text{e} U^\dagger H U P_\text{e}^\dagger.
\label{eq:x2c}
\end{align}
The key feature of all X2C methods is that they reproduce the spectrum of the Dirac equation for 1 particle {\em exactly} (hence, X in X2C). 
For a many-electron system, X2C induces a nonzero error whose magnitude is controlled by 2 factors:
\begin{itemize}
    \item the definition of $U$ (which may be an algorithm, e.g. involve a self-consistent field iteration) and
    \item optional approximations involved in carrying out the unitary transformation.
\end{itemize}
The resulting X2C implementations differ in the details of these two design factors;
see reviews by Liu\cite{KAS:liu:2010:MP} and Wang et. al \cite{KAS:wang:2025:CPR} for detailed overviews.

The simplest and most popular variant of X2C, the \emph{1-electron} (1e) X2C Hamiltonian uses $U$ that block-diagonalizes the 1-body Hamiltonian $h$ but keeps the 2-body counterpart $g$ untransformed:
\begin{align}
H_\text{1eX2C} = P_\text{e} \left( U^\dagger h U + g \right) P_\text{e}^\dagger.
\label{eq:1ex2c}
\end{align}
The 1eX2C Hamiltonian is cheap to construct and robustly accounts for 1e scalar relativistic (SR) and spin-orbit (SO) effects.
However, there are 2 fundamental issues with 1eX2C:
\begin{itemize}
\item {\bf neglect of 2e picture-change (2ePC) effects} due to the use of untransformed $g$ instead of $U^\dagger g U$, and
\item {\bf non-size-intensive picture change and lack of a thermodynamic limit} due to the divergence of the electrostatic  potential of the nuclei entering the 1e Hamiltonian $h$ used to define $U$.
\end{itemize}
The lack of 2ePC in 1eX2C has been shown to significantly impact the accuracy of molecular properties and spectra\cite{KAS:liu:2018:JCP,KAS:knecht:2022:JCP}, thus calling for post-1eX2C treatments.
The latter poses issues for extended systems, such as periodic systems. To avoid this problem, the application of 1eX2C to periodic systems\cite{VRG:abraham:2024:JCTC} must ad hoc replace the nuclear Coulomb potential by its Ewald counterpart. Proper application of X2C to extended systems also requires going beyond 1eX2C to define $U$ using 1-particle operators involving the Coulomb potential of electrons to ensure charge neutrality\cite{VRG:zhao:2016:JCP}.

The straightforward step past 1eX2C is to use the normal-ordered form of $H$, thereby defining the molecular mean-field X2C (mmfX2C)\cite{KAS:liu:2006:JCP,KAS:sikkema:2009:JCP}:\begin{align}
H_\text{mmfX2C} =  E_0 + P_\text{e} \left( U^\dagger f U + w \right) P_\text{e}^\dagger,
\label{eq:mmfx2c}
\end{align}
with $U$ defined to block-diagonalize the 4C Fock matrix. mmfX2C has a well-defined thermodynamic limit and partially accounts for 2ePC; however, the 2-body interactions not included in $f$ are not picture-changed, thereby introducing errors in correlated treatments.
However, mmfX2C is much more expensive than 1eX2C, as it involves the mean-field computation in the bispinor basis.

Due to the largely intra-atomic structure of 2ePC, and especially due to the tightly localized nature of the 2e SO contribution to PC, several compromise X2C methods have been developed that do not involve the unabridged molecular 4C mean-field treatment.
Such methods include:
\begin{itemize}
    \item The Screened-Nuclear-Spin-Orbit (SNSO)\cite{KAS::1962:PRSLA,KAS::1963:PRSLA,KAS:filatov:2013:JCP,KAS:zou:2015:JCP,KAS:jenkins:2019:JCTC,KAS:koulias:2019:JCTC,KAS:hu:2020:JCTC} 
    X2C method corrects 1eX2C for the lack of 2ePC  by heuristic scaling of the 1-electron spin-orbit (1eSO) integrals.\cite{KAS:boettger:2000:PRB}.    
    \item Atomic-mean-field (AMF)-based methods, which correct for 2ePC using appropriately patched atomic densities and potentials obtained from 4C atomic (ensemble) Dirac-Hartree-Fock calculations. These methods include: the AMFI-X2C module\cite{KAS:hess:1996:CPL}, the X2CAMF\cite{KAS:liu:2018:JCP,KAS:zhang:2022:JPCA} scheme, and the recent amfX2C and (extended) eamfX2C schemes\cite{KAS:knecht:2022:JCP}. Henceforth, we shall use the term AMF to refer to amfX2C, eamfX2C, and X2CAMF.
\end{itemize}
Our objective here is to investigate whether it is possible to address 1eX2C's formal deficiencies by defining the operator $U$ and incorporating the 2ePC effects with the help of a model atomic potential. Specifically, we considered the family of potentials that was originally introduced by Lehtola under the moniker SAP (Superposition of Atomic Potentials)\cite{KAS:lehtola:2019:JCTC} for the purpose of constructing robust guess Fock matrices for mean-field calculations. Using the representation of SAP as a combination of contracted s-type Gaussian functions\cite{KAS:lehtola:2020:JCPa}, it is possible to realize the SAP-based X2C approach as a simple modification of the 1eX2C method; the only additional ingredients are derivative three-center two-electron Gaussian AO integrals, which are available in many programs and in open-source Gaussian AO integral libraries.

In \cref{sec:formalism}, we briefly recap the construction of the 1eX2C Hamiltonian, followed by its novel SAP-X2C extension. \cref{sec:technical} provides technical details, and \cref{sec:results} shows the performance assessment of SAP-X2C with respect to the 4C Dirac-Coulomb Hartree Fock (DCHF) and AMF methods for energy and property calculations, as well as a demonstration of the size-intensivity of the SAP-X2C Hamiltonian.

\section{Formalism}\label{sec:formalism}

\subsection{1eX2C}\label{sec:1ex2c}
The restricted kinetic balance (RKB)\cite{VRG:kutzelnigg:1984:IJQC,VRG:stanton:1984:JCP} encodes atomic bispinor $\varphi_\mu$
in terms of scalar AO $\phi_\mu$ as
\begin{align}
  \ket{\varphi_\mu} \equiv
  \begin{pmatrix}
  \ket{\bm{\phi}^\mathrm{L}_\mu} \\
  \ket{\bm{\phi}^\mathrm{S}_\mu}
  \end{pmatrix} = 
  \begin{pmatrix}
  \bm{1} \ket{\phi_\mu} \\
  \bm{\sigma}\cdot\mathbf{p} \ket{\phi_\mu}.
  \end{pmatrix}
\end{align}
Henceforth, L and S will refer to large (upper, electron-like) and small (lower, positron-like) spinors, respectively.

Consider the RKB matrix form of the Dirac equation, also known as the \emph{modified} Dirac equation,\cite{VRG:kutzelnigg:1984:IJQC} for a single particle in potential $V$ expressed in the orthonormal AO (OAO) basis:
\begin{equation}\label{eq:modified-dirac-equation}
  \begin{pmatrix}
    \mathbf{V} & \mathbf{T} \\
    \mathbf{T} & \frac{\mathbf{W}}{4c^2} - \mathbf{T}
  \end{pmatrix}
  \begin{pmatrix}
    \mathbf{C}^\text{L}_{+} & \mathbf{C}^\text{L}_{-} \\
    \mathbf{C}^\text{S}_{+} & \mathbf{C}^\text{S}_{-}
  \end{pmatrix}
  =
  \begin{pmatrix}
    \mathbf{1} & \mathbf{0} \\
    \mathbf{0} & \frac{\mathbf{T}}{2c^2}
  \end{pmatrix}
  \begin{pmatrix}
    \mathbf{C}^\text{L}_{+} & \mathbf{C}^\text{L}_{-} \\
    \mathbf{C}^\text{S}_{+} & \mathbf{C}^\text{S}_{-}
  \end{pmatrix}
  \begin{pmatrix}
    \boldsymbol{\epsilon}_+ & \mathbf{0} \\
    \mathbf{0} & \boldsymbol{\epsilon}_-
  \end{pmatrix},
\end{equation}
where $\boldsymbol{\epsilon}_\pm$ and $\mathbf{C}^\text{L,S}_{\pm}$ are the energies and corresponding OAO coefficients, and $\mathbf{1}$ and $\mathbf{T}$ represent the identity and kinetic energy operators, respectively, in the L orthonormal spinor basis. 
$\mathbf{V}$ and $\mathbf{W}$ are representations of the potential $V$ in the L and S spinor bases, respectively. 

In the 1eX2C method, the transformation $U$ in \cref{eq:1ex2c} is defined to block diagonalize the core Dirac Hamiltonian.
\begin{equation}\label{eq:h-1b-dirac}
\mathcal{H} = 
\begin{pmatrix}
    \mathbf{V} & \mathbf{T} \\
    \mathbf{T} & \frac{\mathbf{W}}{4c^2} - \mathbf{T}
\end{pmatrix}.
\end{equation}
where $V$ contains only the nuclear electrostatic potential. 
Block diagonalization,
\begin{equation}\label{eq:x2c-transformation}
\mathcal{U}^{\dagger} \mathcal{H} \mathcal{U} = 
\begin{pmatrix}
    \mathbf{H}_{++} & \mathbf{0} \\
    \mathbf{0} & \mathbf{H}_{--}
\end{pmatrix},
\end{equation}
is achieved by
\begin{equation}\label{eq:unitary-block-diagonalizer}
\mathcal{U} = 
\begin{pmatrix}
   \mathbf{R} & -\mathbf{X}^\dagger \mathbf{R}^\dagger \\
    \mathbf{XR} & \mathbf{R}^\dagger
\end{pmatrix},
\end{equation}
where $\mathbf{X}$ is obtained by solving the linear system of equations,
\begin{equation}\label{eq:x-mat}
     \quad  \mathbf{C}^\text{L}_{+} \mathbf{X} = \mathbf{C}^\text{S}_{+}
\end{equation}
and
\begin{equation}\label{eq:r-mat}
    \mathbf{R} \equiv \, \frac{1}{\sqrt{\mathbf{I} + \mathbf{X}^\dagger \mathbf{X}}}.
\end{equation}

The 1eX2C Hamiltonian,
\begin{equation}\label{eq:h-1eX2C}
    \mathbf{H}^{\text{1eX2C}} \equiv \mathbf{H}_{++} = \mathbf{R}^\dagger \left[ \mathbf{V} + \mathbf{TX} + \mathbf{X}^\dagger \mathbf{T} + \mathbf{X}^\dagger (\frac{\mathbf{W}}{4c^2} -\mathbf{T}) \mathbf{X} \right] \mathbf{R}.
\end{equation}
has a spectrum identical to the positive spectrum of $\mathcal{H}$. 
Details of practical implementation are described, for example, in Ref. \citenum{KAS:peng:2013:JCP}.
However, the precision of the resulting 1eX2C Hamiltonian can be sensitive to numerical errors that occur in the course of diagonalizing matrices $\mathbf{S}$ and  $\mathbf{T}$ whose condition numbers routinely exceed $10^8$; the X2C implementation in \code{MPQC} uses SVD instead of the eigensolver if the condition number exceeds $10^8$.

\subsection{SAP-X2C}\label{sec:sap}

The original definition of Superposition of Atomic Potentials\cite{KAS:lehtola:2019:JCTC} expressed the contribution from atom $A$ as:
\begin{align}\label{eq:v-sap-r}
    V^{\text{SAP}}_A(\mathbf{r}) = & -\frac{Z_A^{\text{eff}}(r_A)}{r_A},
\end{align}
where $r_A = |\mathbf{r} - \mathbf{R}_A|$, and the effective nuclear charge function $Z^{\text{eff}}_A$ is defined on a fixed radial grid by fitting to the reference potentials. Although the atomic potentials in SAP are local, they inherit the local description of exchange and correlation effects from the reference potential used for the fitting.
Ref. \citenum{KAS:lehtola:2020:JCPa} replaced the original prescription by a combination of the bare nuclear potential and the contribution due to the screening of the bare nucleus by other electrons:
\begin{align} \label{eq:v-sap-r2}
    V_A^{\text{SAP}}(\mathbf{r}) = & -  \frac{Z_A}{r_A} + \int \mathrm{d}\mathbf{r}' \, \frac{\theta_A(\mathbf{r}')}{|\mathbf{r} - \mathbf{r}'|}, \\
    \theta_A(\mathbf{r}) \equiv & \sum_k c_k  \exp(-\alpha_k |\mathbf{r} - \mathbf{R}_A|^2 ).
\end{align}
Matrix elements over SAP in this form can be computed straightforwardly in any electronic structure package:
\begin{align}\label{eq:v-sap}
V^{\text{SAP}}_{\mu\nu} &= V_{\mu\nu} + V^{\text{e}}_{\mu\nu}, \\
V^{\text{e}}_{\mu\nu} &= - \sum_{A} (\theta_{A}| \phi_{\mu} \phi_{\nu})
\end{align}
where $V_{\mu\nu}$ are the matrix elements of the nuclear Coulomb potential, and 
the screening contribution $V^{\text{e}}_{\mu\nu}$ consists of three-center two-electron integrals over contracted s-type Gaussian functions $\theta_{A}$:
\begin{align}
  (\theta_{A}| \phi_{\mu} \phi_{\nu}) \equiv \iint \mathrm{d}\mathbf{r} \, \mathrm{d}\mathbf{r}' \,\theta_{A}(\mathbf{r}) |\mathbf{r} - \mathbf{r}'|^{-1} \phi_{\mu}(\mathbf{r}') \phi_{\nu}(\mathbf{r}').
\end{align}
The S spinor integrals over SAP are expressed similarly:
\begin{align}
W^{\text{SAP}}_{\mu\nu} &= W_{\mu\nu} + W^{\text{e}}_{\mu\nu} \\
\label{eq:w-sap-e-integral}
W^{\text{e}}_{\mu\nu} &\equiv -\sum_{A} (\theta_{A}| [(\boldsymbol{\sigma} \cdot \mathbf{p})\phi_{\mu}] [(\boldsymbol{\sigma} \cdot \mathbf{p})\phi_{\nu}])
\end{align}
$W^{e}_{\mu\nu}$ can be readily evaluated in terms of second-order geometrical derivatives of three-center two-electron integrals.
Such integrals are available in many molecular electronic structure packages and in standalone Gaussian AO integral libraries such as \code{Libint}\cite{Libint2,libint-2.10.0} and \code{Libcint}\cite{VRG:sun:2015:JCC,VRG:sun:2024:JCP}.
However, since only some mixed second-order derivatives are needed to obtain \cref{eq:w-sap-e-integral}, it is more efficient to evaluate such integrals using dedicated kernels.

Note that \cref{eq:v-sap-r,eq:v-sap-r2} are both defined\cite{KAS:lehtola:2019:JCTC,KAS:lehtola:2020:JCPa} to ensure superpolynomial decay, namely for any $p\geq 0$
\begin{align}
\label{eq:v-sap-asymptotics}
\lim_{r_A \to \infty} r_A^p V_A^{\text{SAP}}(\mathbf{r}) & = 0.
\end{align}
If condition \cref{eq:v-sap-asymptotics} is valid for $p\geq d$ SAP will be absolutely convergent for a $d$-dimensional lattice of atoms. The superpolynomial decay thus helps guarantee SAP's absolute convergence and efficient evaluation (due to its short-ranged character). This is critical to the applicability of SAP to large systems in general and will be 
important in \cref{sec:size-intensivity-test}.

Substituting $V\to V^\text{SAP},W\to W^\text{SAP}$ in the modified Dirac Hamiltonian \cref{eq:h-1b-dirac} produces the corresponding SAP-Dirac Hamiltonian:
\begin{equation}\label{eq:h-1b-sap-dirac}
    \mathcal{H}^{{\text{SAP}}} = 
    \begin{pmatrix}
    \mathbf{V}^{\text{SAP}} & \mathbf{T} \\
    \mathbf{T} & \frac{\mathbf{W}^{\text{SAP}}}{4c^2} - \mathbf{T}
    \end{pmatrix}
    .
\end{equation}
Its X2C block diagonalization produces the SAP counterpart of the 1eX2C core Hamiltonian, which includes the X2C-transformed screening contribution $V^\text{e}$ to the SAP potential \cref{eq:v-sap}. To make the transition from 1eX2C to SAP-X2C as simple as possible, it is convenient to define the SAP-X2C \emph{core} Hamiltonian as
\begin{equation}\label{eq:h-x2c1e-sap}
    \mathbf{H}^{\text{SAP-X2C}} = \mathbf{R}'^\dagger \left[ \mathbf{V}^{\text{SAP}} + \mathbf{TX}' + \mathbf{X}'^\dagger \mathbf{T} + \mathbf{X}'^\dagger (\frac{\mathbf{W}^{\text{SAP}}}{4c^2} -\mathbf{T}) \mathbf{X}' \right] \mathbf{R}'  - \mathbf{V}^{\text{e}},
\end{equation}
where the matrices $\mathbf{X}'$ and $\mathbf{R}'$ are obtained following the recipe from \cref{eq:modified-dirac-equation,eq:x-mat,eq:r-mat}, the screening contribution to SAP, $\mathbf{V}^{\text{e}}$, is subtracted at the end to avoid double counting of the 2e contribution in subsequent calculations. The SAP-X2C core Hamiltonian $\mathbf{H}^{\text{SAP-X2C}}$ can then be used in place of its counterpart $\mathbf{H}^{\text{1eX2C}}$ (\cref{eq:h-1eX2C}) in mean-field and correlated calculations.

Here, we would like to point out that ZORA(MP, i.e., \emph{model-potential}) by van Wüllen\cite{KAS:vanwullen:1998:JCP} does share some connections with SAP-X2C; however,, the model-potential constructed in Ref.\cite{KAS:vanwullen:1998:JCP} does not account 2ePC effects.

\section{Technical details}\label{sec:technical}

The 1eX2C and SAP-X2C Hamiltonians were implemented in the Massively Parallel Quantum Chemistry (\code{MPQC}) package\cite{VRG:peng:2020:JCP}. The SAP ``basis'' \code{sap\_grasp\_large} from Basis Set Exchange (BSE)\cite{VRG:pritchard:2019:JCIM} is used for all SAP-X2C calculations in this work, as it is optimized with numerical Dirac-Coulomb Hartree Fock\cite{KAS:lehtola:2020:JCPa}. Note that the normalization convention for the SAP ``basis'' is not the same as the convention for Gaussian AO basis sets, see Ref.\citenum{KAS:lehtola:2020:JCPa}; the SAP definitions were incorporated into the \code{Libint} library's public 2.10.0 release.\cite{libint-2.10.0}

The molecular geometries used for the comparison of HF energies in \cref{fig:error-in-hf-energies} were optimized using \code{Psi4}\cite{VRG:smith:2020:JCP} with PBE0\cite{KAS:adamo:1999:JCP} and def2-TZVP\cite{KAS:weigend:2005:PCCP} basis set. The molecular geometry of the Og$_2$ molecule used for spinor energy assessment was taken from Ref.\citenum{KAS:knecht:2022:JCP}. The dyall-ae3z\cite{VRG:gomes:2010:TCA,KAS:dyall:2012:TCA,KAS:dyall:2023:} basis set was used for all calculations in this work, except for the size-intensivity test, where dyall-2zp was used for efficiency. The physical constants defined in CODATA-2022\cite{KAS:mohr:2025:RMP} were used throughout, including the speed of light of 137.03599917697 $a_0 E_h/\hbar$ obtained from the fine structure constant.
The 4C-DCHF, amfX2C and eamfX2C calculations were performed using \code{DIRAC}\cite{KAS:saue:2020:JCP}. The prerequisite atomic calculations needed for amf and eamfX2C were done using \emph{the average-of-configuration} (AOC)\cite{KAS:zerner:1989:IJQC} mmfX2C with ground state electronic configurations taken from NIST Atomic Spectra Database\cite{VRG:kramida:1999:}.
The X2CAMF calculations were performed using \code{socutils}\cite{socutils}. Since X2CAMF includes only 2eSO-PC and not 2eSR-PC, we are comparing against X2CAMF only for spin-orbit splitting energies, equilibrium bond distances, and harmonic vibrational frequencies.

Harmonic vibrational frequencies of coinage and halogen dimers were computed by 4th-order polynomial fit of the potential energies evaluated at $R=\{R_{\text{eq}},R_{\text{eq}} \pm 0.005 \text{\AA}, R_{\text{eq}} \pm 0.01 \text{\AA}\}$. The atomic masses used in this work are listed in \cref{tab:atomic-masses}.

\begin{table}[ht!]
    \centering
    \begin{tabular}{c|r}
    \hline\hline
    Isotope & Mass (a.u.) \\ 
    \hline
    $^{63}$Cu   &  62.9295989 \\
    $^{107}$Ag   &  106.905092 \\
    $^{197}$Au   &  196.966543 \\
    $^{79}$Br   &  78.9183361 \\
    $^{127}$I    &  126.904473 \\
    $^{210}$At   &  209.987126 \\
    \hline\hline
    \end{tabular}
    \caption{Atomic masses used for the harmonic vibrational frequency computations.}
    \label{tab:atomic-masses}
\end{table}

For the size-intensivity test in \cref{sec:size-intensivity-test} used fragments of Xe crystal of increasing size; the \code{.xyz} files for these fragments are available in the Supporting Information.

\section{Results}\label{sec:results}

\subsection{Absolute Energies}\label{sec:mf-energies}

\begin{figure}[ht!]
    \centering
    \includegraphics[width=0.5\textwidth]{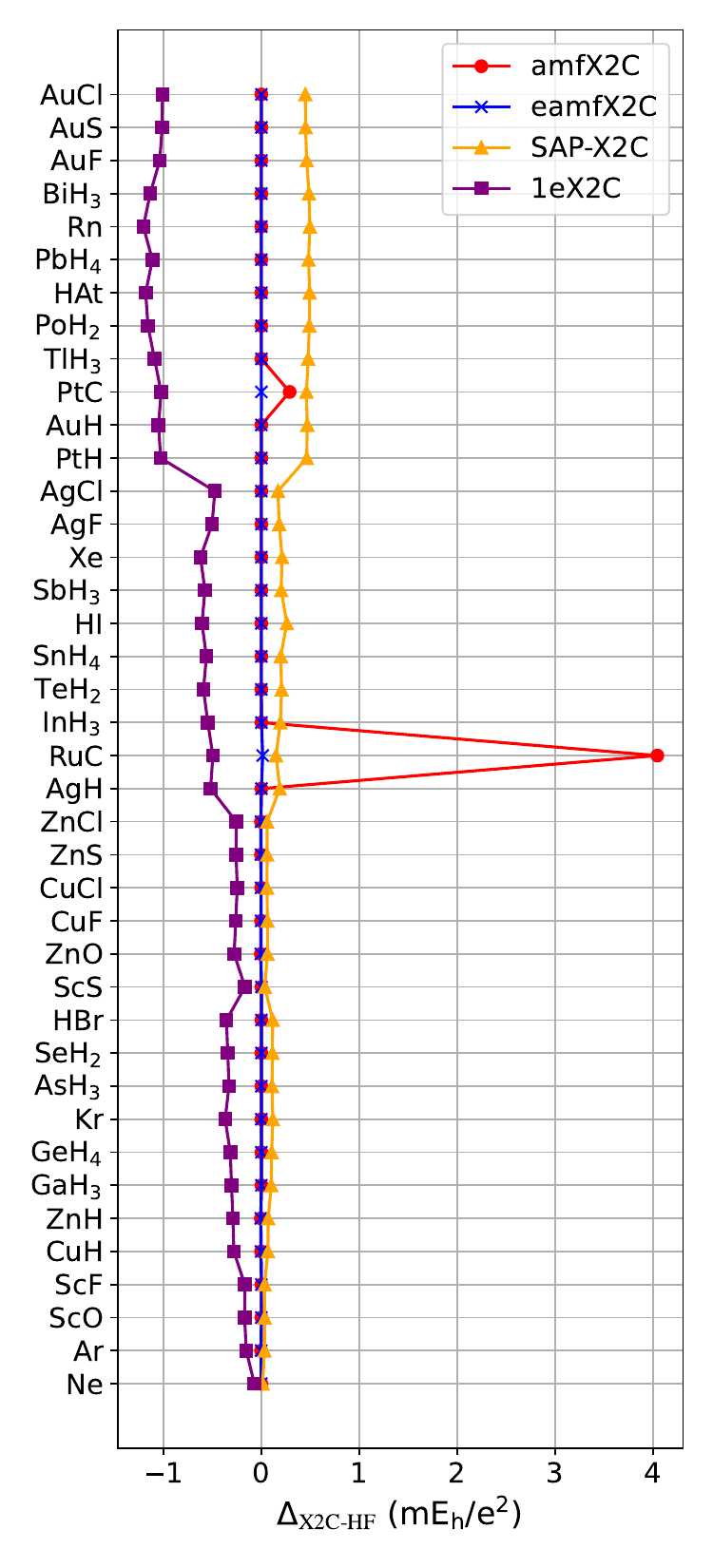}
    \caption{Errors of $Z$-adjusted X2C-HF energies with respect to 4C-DCHF. $\Delta_{\mathrm{X2C-HF}} = (E_{\mathrm{X2C-HF}} - E_{\mathrm{4C-DCHF}})/ \sum_A Z_A^2$, where $\sum_A Z_A^2$ is sum of squares of atomic numbers of all atoms in the system.}
    \label{fig:error-in-hf-energies}
\end{figure}

\cref{fig:error-in-hf-energies} shows errors in the total HF energies obtained with a variety of X2C approaches for a diverse set of small molecules containing light and heavy elements. Due to the lack of 2ePC, the 1eX2C errors clearly increase with $Z$. The amfX2C approach is highly accurate for most of the molecules presented, except for two outliers: RuC and PtC. eamfX2C, on the other hand, is highly accurate throughout; however, it is more expensive than amfX2C, as it involves a singular construction of a molecular 4C Fock matrix. SAP-X2C is a significant improvement over the 1eX2C and shows remarkably small errors even for heavy-element-containing systems.

The accuracy of HF spinor energies (\cref{tab:spinor-energies-og2}) was assessed using the same noble gas dimer Og$_2$ that was previously used by Knecht et al.\cite{KAS:knecht:2022:JCP}. However, unlike Ref. \citenum{KAS:knecht:2022:JCP}, we used the dyall-ae3z AO basis set in this work. Errors in spinor energies are significantly smaller with SAP-X2C than with 1eX2C; this is especially pronounced in the valence part of Og$_2$. Note that the remarkable accuracy of AMF models is somewhat fortuitous due to the largely atom-like character of the electronic structure in this system (modulo the symmetry lowering).

\cref{tab:spin-orbit-splitting} illustrates spin-orbit splittings of several core subshells of the Rn and Og atoms. SAP-X2C yet again is more accurate than 1eX2C. In total HF energies, spinor energies, and splitting energies, amfX2C and eamfX2C outperform SAP-X2C in terms of accuracy.
However, SAP-X2C does a fairly good job considering that it does not require any prerequisite atomic calculations, which makes it significantly cheaper and more black-box than the AMF methods and a viable option for large scale calculations. 

\begin{table}[ht!]
  \centering
  \sisetup{
    scientific-notation = true,
    table-number-alignment = center,
    print-zero-exponent = true,
    table-align-exponent = true,
    input-symbols = {\dots},     % allow \dots in S columns
    table-space-text-pre = {\dots}, % reserve width if needed
    detect-weight = true
  }
  \resizebox{\textwidth}{!}{%
  \begin{tabular}{r|
                  r|
                  S[table-format=2.2e+1] |
                  S[table-format=2.2e+1] |
                  S[table-format=2.2e+1] |
                  S[table-format=2.2e+1] |
                  S[table-format=2.2e+1] }
    \hline\hline
     $\epsilon_i$ & 4C-DCHF ($E_{h}$) & {amfX2C} & {eamfX2C} & {SAP-X2C} & {1eX2C} & {sf-1eX2C}\\
    \hline
    $\epsilon_{1-2}$ & -8272.084 & {\bfseries \num{3.64e-02}} & 3.90e-02 & 7.25e+03 & 2.38e+04 & 8.12e+03 \\
    $\epsilon_{3-4}$ & -1738.989 & {\bfseries \num{3.55e-02}} & 3.81e-02 & 2.39e+03 & 5.11e+03 & 3.62e+02 \\
    $\epsilon_{5-6}$ & -1686.480 & {\bfseries \num{3.07e-02}} & 3.33e-02 & -1.51e+03 & -7.24e+03 & 4.23e+05 \\
    $\epsilon_{7-10}$ & -1137.975 & {\bfseries \num{3.01e-02}} & 3.36e-02 & 4.91e+02 & 4.04e+03 & -1.26e+05 \\
    $\epsilon_{11-12}$ & -476.181 & {\bfseries \num{3.04e-02}} & 3.29e-02 & 6.05e+02 & 1.21e+03 & -3.50e+02 \\
    $\epsilon_{13-14}$ & -453.055 & {\bfseries \num{2.89e-02}} & 3.15e-02 & -3.60e+02 & -1.74e+03 & 1.03e+05 \\
    $\epsilon_{15-18}$ & -318.144 & {\bfseries \num{2.71e-02}} & 3.02e-02 & 1.36e+02 & 1.04e+03 & -3.21e+04 \\
    $\epsilon_{19-22}$ & -286.472 & {\bfseries \num{2.81e-02}} & 3.22e-02 & -1.43e+02 & -1.38e+03 & 1.22e+04 \\
    $\epsilon_{23-28}$ & -265.518 & {\bfseries \num{2.67e-02}} & 3.08e-02 & 6.77e+01 & 1.00e+03 & -8.71e+03 \\
    $\epsilon_{29-30}$ & -142.437 & {\bfseries \num{2.54e-02}} & 2.79e-02 & 1.75e+02 & 3.35e+02 & -2.02e+02 \\
    $\epsilon_{31-32}$ & -131.405 & {\bfseries \num{2.49e-02}} & 2.75e-02 & -1.09e+02 & -5.36e+02 & 2.99e+04 \\
    $\epsilon_{33-36}$ & -91.953 & {\bfseries \num{2.60e-02}} & 2.61e-02 & 4.06e+01 & 3.01e+02 & -9.60e+03 \\
    $\epsilon_{37-40}$ & -76.202 & {\bfseries \num{2.33e-02}} & 2.74e-02 & -4.16e+01 & -4.17e+02 & 3.38e+03 \\
    $\epsilon_{41-46}$ & -70.293 & 2.70e-02 & {\bfseries \num{2.62e-02}} & 1.89e+01 & 2.79e+02 & -2.53e+03 \\
    $\epsilon_{47-52}$ & -49.743 & {\bfseries \num{2.03e-02}} & 2.68e-02 & -2.49e+01 & -3.52e+02 & 7.78e+02 \\
    $\epsilon_{53-60}$ & -47.995 & {\bfseries \num{2.56e-02}} & 2.66e-02 & 1.00e+01 & 2.42e+02 & -9.69e+02 \\
    $\dots$ & $\dots$ & $\dots$ & $\dots$ & $\dots$ & $\dots$ & $\dots$ \\
    $\epsilon_{111}$ & -1.323 & {\bfseries \num{1.55e-02}} & 1.78e-02 & 2.57e+00 & 2.89e+00 & 1.50e+01 \\
    $\epsilon_{112}$ & -1.322 & {\bfseries \num{1.51e-02}} & 1.78e-02 & 2.58e+00 & 2.91e+00 & 1.50e+01 \\
    $\epsilon_{113}$ & -0.746 & {\bfseries \num{1.17e-02}} & 1.51e-02 & -1.70e+00 & -8.60e+00 & 3.34e+02 \\
    $\epsilon_{114}$ & -0.743 & {\bfseries \num{1.33e-02}} & 1.50e-02 & -1.72e+00 & -8.67e+00 & 3.49e+02 \\
    $\epsilon_{115}$ & -0.323 & 8.46e-03 & {\bfseries \num{6.94e-03}} & 9.54e-02 & 1.11e+00 & -7.16e+01 \\
    $\epsilon_{116}$ & -0.310 & 9.32e-03 & {\bfseries \num{6.84e-03}} & 1.03e-01 & 1.21e+00 & -7.79e+01 \\
    $\epsilon_{117}$ & -0.298 & 4.68e-03 & {\bfseries \num{4.62e-03}} & 1.15e-01 & 1.34e+00 & -8.96e+01 \\
    $\epsilon_{118}$ & -0.287 & -2.43e-03 & {\bfseries \num{1.06e-02}} & 1.17e-01 & 1.39e+00 & -8.50e+01 \\
    \hline\hline
    \end{tabular}
    }
    \caption{4C-DCHF spinor energies and the corresponding errors of X2C-HF spinor energies in the Og$_2$ molecule ($R_{\text{eq}}$ = 4.329 \AA)\cite{KAS:knecht:2022:JCP}. The smallest X2C error for each spinor is shown in bold.} 
    \label{tab:spinor-energies-og2}
\end{table}

\begin{table}[ht!]
  \centering
  \sisetup{
    scientific-notation = true,
    table-number-alignment = center,
    print-zero-exponent = true,
    table-align-exponent = true,
    detect-weight = true
  }
  \resizebox{\textwidth}{!}{%
    \begin{tabular}{r|
                    r|
                    r|
                    S[table-format=2.2e+1] |
                    S[table-format=2.2e+1] |
                    S[table-format=2.2e+1] |
                    S[table-format=2.2e+1] |
                    S[table-format=2.2e+1]}
      \hline\hline
      Atom & Subshell & {4C-DCHF} & {amfX2C} & {eamfX2C} & {X2CAMF} & {SAP-X2C} & {1eX2C} \\
      \hline
      Rn   & 2p & 2755.681 & {\bfseries \num{-3.67e-06}} & {\bfseries \num{-3.67e-06}} & -1.93e0 & 1.73e1 & 7.93e1 \\
           & 3p & 630.632 & {\bfseries \num{-7.10e-06}} & -7.13e-06 & -6.83e-01 & 3.99e0 & 1.90e1 \\
           & 3d & 130.823 & {\bfseries \num{-3.13e-05}} & {\bfseries \num{-3.13e-05}} & -2.02e-02 & 1.96e0 & 1.93e1 \\
           & 4f & 7.229 & {\bfseries \num{-3.56e-06}} & {\bfseries \num{-3.56e-06}} & 1.30e-03 & 2.33e-01 & 3.65e0 \\
      Og   & 2p & 14925.605 & {\bfseries \num{-1.24e-05}} & {\bfseries \num{-1.24e-05}} & -1.92e1 & 5.45e1 & 3.07e2 \\
           & 3p & 3671.111 & {\bfseries \num{-2.56e-05}} & {\bfseries \num{-2.56e-05}} & -8.46e0 & 1.35e1 & 7.55e1 \\
           & 3d & 570.18 & {\bfseries \num{-3.67e-05}} & {\bfseries \num{-3.67e-05}} & -3.68e-02 & 5.74e0 & 6.48e1 \\
           & 4f & 47.543 & -4.24e-06 & {\bfseries \num{-4.22e-06}} & 6.19e-03 & 9.51e-01 & 1.62e1 \\
      \hline\hline
    \end{tabular}%
  }
  \caption{Errors of X2C-HF spin-orbit splittings (eV) relative to 4C-DCHF for inner subshells of heavy noble gas atoms. The smallest X2C error for each subshell is shown in bold.}
  \label{tab:spin-orbit-splitting}
\end{table}

\subsection{Molecular Properties}\label{sec:mf-properties}

\begin{figure}[ht!]
    \centering
    % First row: Coinage
    \begin{subfigure}[b]{0.328\textwidth}
        \includegraphics[width=\textwidth]{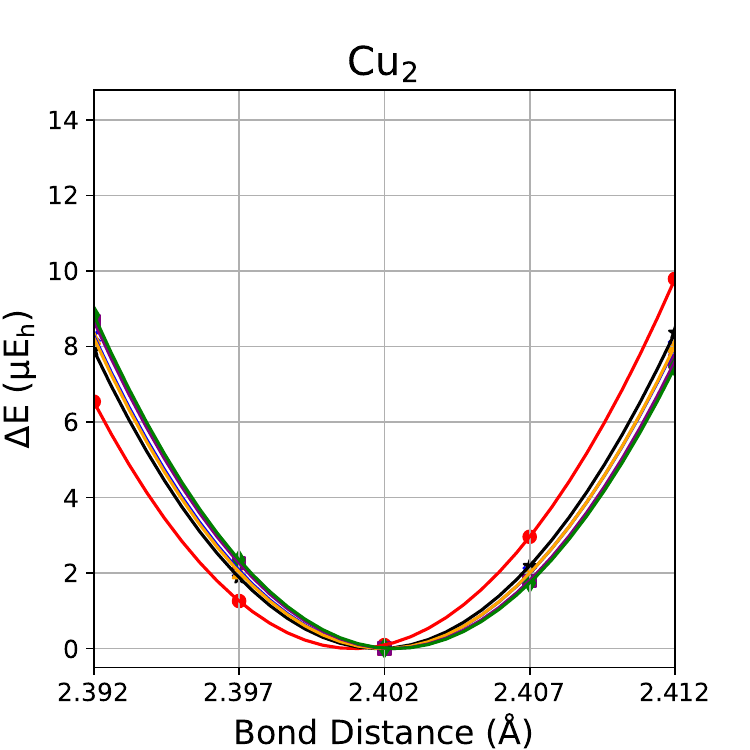}
        \caption{}
        \label{fig:pes_Cu2}
    \end{subfigure}
    \hfill
    \begin{subfigure}[b]{0.328\textwidth}
        \includegraphics[width=\textwidth]{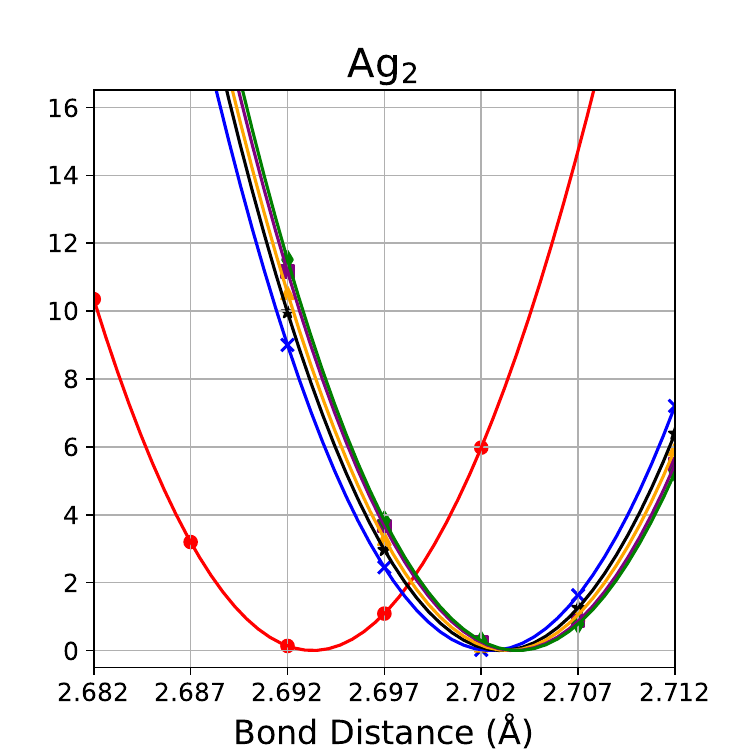}
        \caption{}
        \label{fig:pes_Ag2}
    \end{subfigure}
    \hfill
    \begin{subfigure}[b]{0.328\textwidth}
        \includegraphics[width=\textwidth]{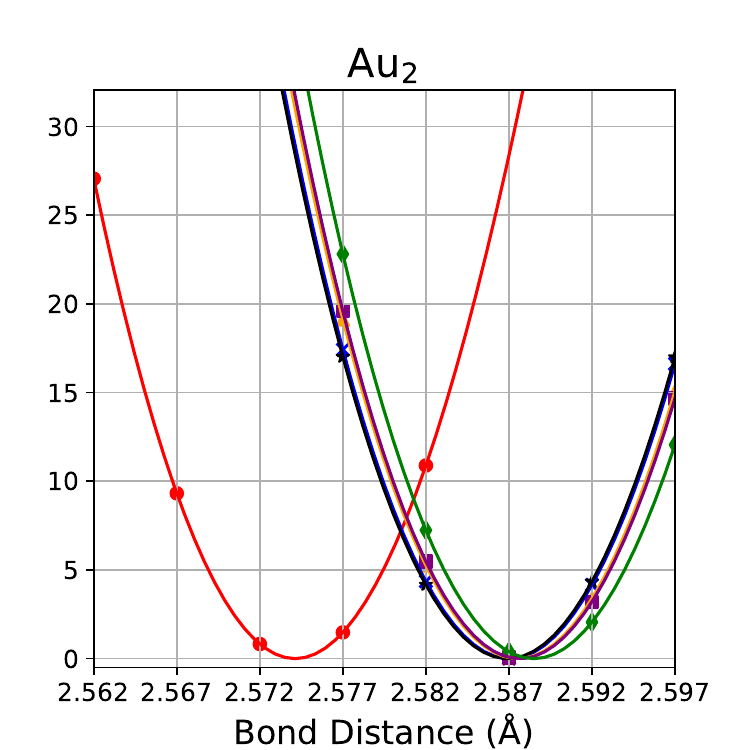}
        \caption{}
        \label{fig:pes_Au2}
    \end{subfigure}
    
    \vspace{0.3cm}
    
    % Second row: Halogens
    \begin{subfigure}[b]{0.328\textwidth}
        \includegraphics[width=\textwidth]{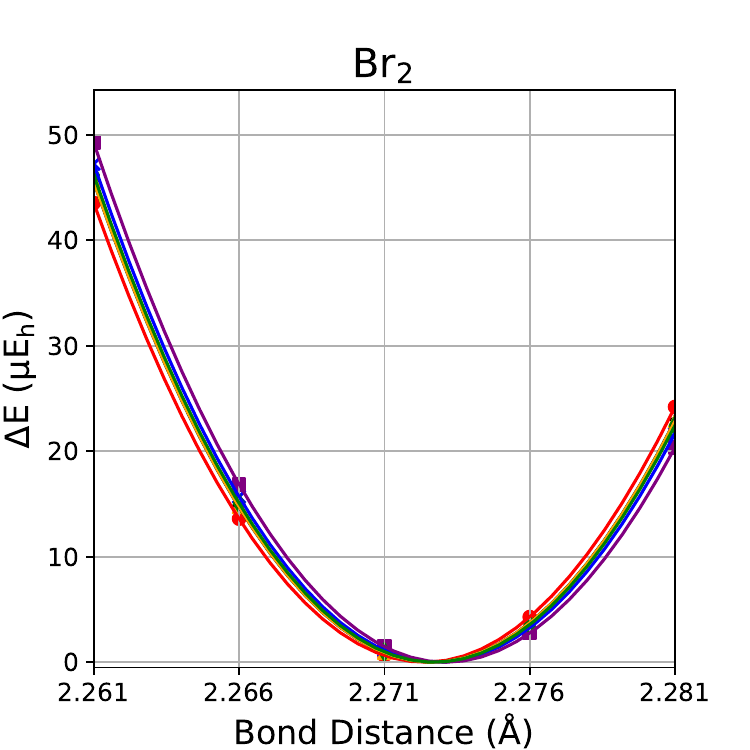}
        \caption{}
        \label{fig:pes_Br2}
    \end{subfigure}
    \hfill
    \begin{subfigure}[b]{0.328\textwidth}
        \includegraphics[width=\textwidth]{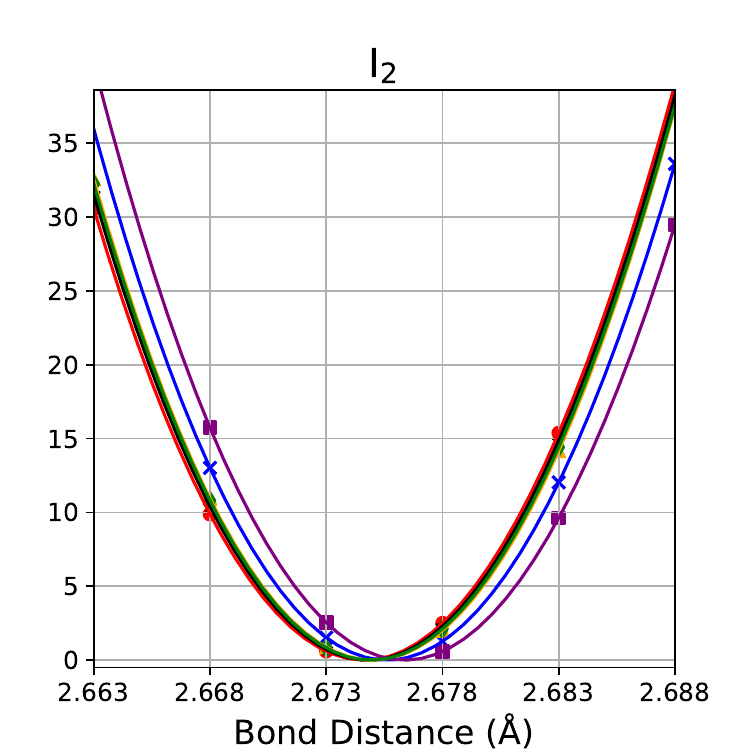}
        \caption{}
        \label{fig:pes_I2}
    \end{subfigure}
    \hfill
    \begin{subfigure}[b]{0.328\textwidth}
        \includegraphics[width=\textwidth]{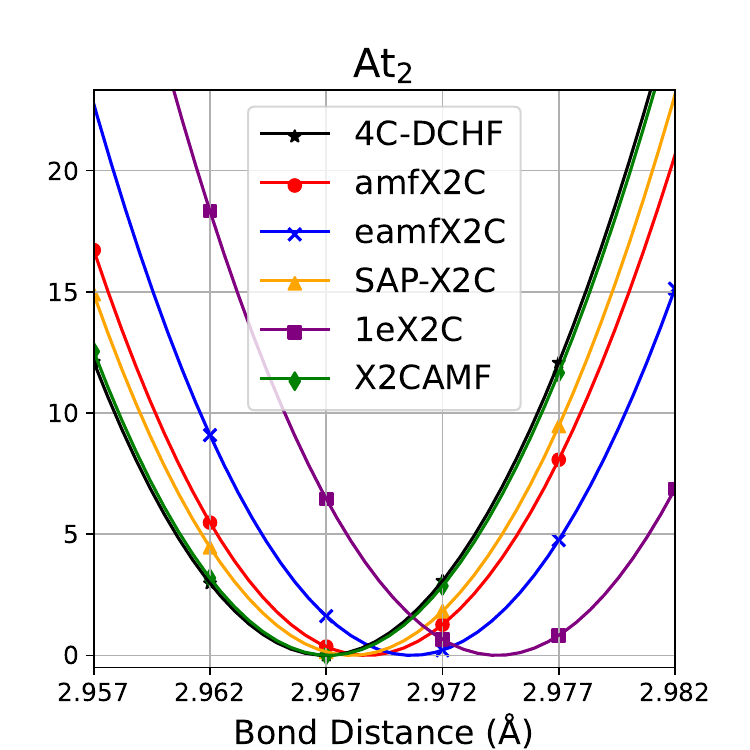}
        \caption{}
        \label{fig:pes_At2}
    \end{subfigure}

    \caption{Aligned potential energy surfaces (PESs) of coinage dimers (\ref{fig:pes_Cu2},\ref{fig:pes_Ag2},\ref{fig:pes_Au2}) and halogen dimers (\ref{fig:pes_Br2},\ref{fig:pes_I2},\ref{fig:pes_At2}) with 4C-DCHF and  X2C-HF methods. $\Delta E \equiv E(R) - E(R_\text{eq})$.}
    \label{fig:dimers_PESs}
\end{figure}

Next, we assessed the accuracy of molecular potential energy surfaces (PES) obtained with SAP-X2C against the 4C-DCHF and other X2C variants; fitted PES are illustrated in \cref{fig:dimers_PESs}, with the corresponding equilibrium bond distances and harmonic vibrational frequencies shown in
\cref{tab:bondlengths,tab:harmonic-frequencies}, respectively.

Unlike the absolute energies and spin-orbit splittings, where SAP-X2C (while far superior to 1eX2C) never approached the performance of AMF-based X2C variants,
SAP-X2C bond distances and harmonic vibrational frequencies were the most accurate among all X2C variants for 3 and 2 molecules, respectively (out of 6). Surprisingly, amfX2C produced large errors for bond distances of Ag$_2$ and Au$_2$ ($~10~\text{m}\AA$) and for the vibrational frequency of Au$_2$ ($~5~\text{cm}^{-1}$). While for most systems all X2C variants reproduce the reference bond distances to better than 1 milliangstrom, it is also intriguing to see the relatively wide distribution of the X2C equilibrium bond distances for At$_2$ (\cref{fig:pes_At2}). The observed excellent performance of SAP-X2C for equilibrium geometry and vibrational frequency computation makes it a prime candidate for inexpensive and accurate navigation of relativistic molecular PES. Note that the differences in bond distances and vibrational frequencies resulting from different X2C Hamiltonians are relatively small compared to the residual errors of the electronic structure model itsel. This is illustrated in \cref{tab:dft-bonddist-and-freqs} that juxtaposes the equilibrium  bond distances and harmonic vibrational frequencies obtained with with SAP-X2C and 1eX2C Hamiltonians using KS DFT (the hybrid PBE0 functional was used). The differences are relatively small relative to the HF-KS difference, but nevertheless becomes non-negligible for the heaviest systems.

\begin{table}[ht!]
  \centering
  \sisetup{
    scientific-notation = true,
    table-number-alignment = center,
    print-zero-exponent = true,
    table-align-exponent = true,
    detect-weight          = true
  }
  \resizebox{\textwidth}{!}{%
  \begin{tabular}{r|
                  r|
                  S[table-format=2.2e+1] |
                  S[table-format=2.2e+1] |
                  S[table-format=2.2e+1] |
                  S[table-format=2.2e+1] |
                  S[table-format=2.2e+1]}
    \hline\hline
    Molecules & 4C-DCHF & {amfX2C} & {eamfX2C} & {X2CAMF} & {SAP-X2C} & {1eX2C} \\
    \hline
    Cu$_2$ & 2.4018 & -8.61e-04  & 2.36e-04 & 5.67e-04 & {\bfseries \num{2.09e-04}} & 4.75e-04 \\
    Ag$_2$ & 2.7030 & -9.74e-03 & -5.48e-04 & 8.36e-04 & {\bfseries \num{3.32e-04}} & 6.81e-04 \\
    Au$_2$ & 2.5869 & -1.28e-02 & {\bfseries \num{1.12e-04}} & 1.58e-03 & 6.07e-04 & 7.19e-04 \\
    Br$_2$ & 2.2726 & -2.45e-04 & 2.21e-04 & 8.21e-05 & {\bfseries \num{8.53e-06}} & 4.76e-04 \\
    I$_2$ & 2.6748 & -1.60e-04 & 8.14e-04 & {\bfseries \num{1.26e-04}} & 1.91e-04 & 1.58e-03 \\
    At$_2$ & 2.9669 & 1.79e-03 & 3.73e-03 & {\bfseries \num{1.74e-04}} & 1.15e-03 & 7.41e-03 \\
    \hline\hline
    \end{tabular}
    }
    \caption{The 4C-DCHF equilibrium bond distances and the corresponding X2C-HF errors (\AA) for coinage metal and heavy halogen dimers. The smallest X2C error for each molecule is shown in bold.}
    \label{tab:bondlengths}
\end{table}

\begin{table}[ht!]
  \centering
  \sisetup{
    scientific-notation = true,
    table-number-alignment = center,
    print-zero-exponent = true,
    table-align-exponent = true,
    detect-weight  = true
  }
  \resizebox{\textwidth}{!}{%
  \begin{tabular}{r|
                  r|
                  S[table-format=2.2e+1] |
                  S[table-format=2.2e+1] |
                  S[table-format=2.2e+1] |
                  S[table-format=2.2e+1] |
                  S[table-format=2.2e+1]}
    \hline\hline
    Molecules & {4C-DCHF} & {amfX2C} & {eamfX2C} & {X2CAMF} & {SAP-X2C} & {1eX2C} \\
    \hline
    Cu$_2$ & 195.60 & -3.48e-01 & {\bfseries \num{-3.29e-02}} & -1.58e-01 & -1.34e+00 & -5.54e-01 \\
    Ag$_2$ & 149.37 & -6.57e-01 & 4.59e-02 & -2.09e-01 & {\bfseries \num{1.05e-02}} & 1.63e-01 \\
    Au$_2$ & 159.89 & 4.99e+00 & {\bfseries \num{5.97e-02}} & -3.82e-01 & 3.35e-01 & 7.62e-02 \\
    Br$_2$ & 352.27 & -3.14e-01 & -2.10e-01 & -2.96e-02 & {\bfseries \num{-8.21e-03}} & -2.30e-01 \\
    I$_2$ & 228.18 & -4.37e-01 & -3.83e-01 & {\bfseries \num{-1.68e-02}} & -7.10e-01 & -8.69e-01 \\
    At$_2$ & 130.51 & -6.33e-01 & -7.03e-01 & {\bfseries \num{-1.45e-02}} & -7.75e-01 & -1.40e+00 \\
    \hline\hline
    \end{tabular}
    }
    \caption{The 4C-DCHF harmonic vibrational frequencies and the corresponding X2C-HF errors (cm$^{-1}$) for coinage metal and heavy halogen dimers. The smallest X2C error for each molecule is shown in bold.}
    \label{tab:harmonic-frequencies}
\end{table}

\begin{table}[ht!]
  \centering
  \sisetup{
    scientific-notation = true,
    table-number-alignment = center,
    print-zero-exponent = true,
    table-align-exponent = true,
    detect-weight = true
  }
  % \resizebox{\textwidth}{!}{%
  \begin{tabular}{l |
                  l |
                  l |
                  l |
                  l |
                  l |
                  l }
    \hline\hline
    & \multicolumn{3}{c|}{Equilibrium Bond Distance (\AA)} &
      \multicolumn{3}{c}{Harmonic Vibrational Frequency (cm$^{-1}$)} \\
    \cline{2-7}
    Molecule & {1eX2C} & {SAP-X2C} & {Experiment} &
               {1eX2C} & {SAP-X2C} & {Experiment} \\
    \hline
    Cu$_2$ &2.2358&2.2356&2.218\cite{KAS:bornhauser:2020:JCP}&259.13&259.21& 266.49\cite{KAS:bornhauser:2020:JCP}\\
    Ag$_2$ &2.5685&2.5682&2.530\cite{KAS:beutel:1993:JCP}&181.97&181.94& 192.4\cite{kleman1955_Ag_omegae}\\
    Au$_2$ &2.4997&2.4994&2.472\cite{KAS:morse:1986:CR,KAS:beutel:1993:JCP,KAS:simard:1990:JoMS,KAS:ames:1967:TFS}&186.50&186.24&190.9\cite{KAS:morse:1986:CR,KAS:beutel:1993:JCP,KAS:simard:1990:JoMS,KAS:ames:1967:TFS}\\
    Br$_2$ &2.2805&2.28&2.281\cite{KAS:holzer:1970:JCP,KAS:baierl:1975:AS}&332.59&333.04&325.32\cite{KAS:holzer:1970:JCP,KAS:baierl:1975:AS}\\
    I$_2$  &2.6742&2.6725&2.665\cite{KAS:howard:1974:JRS}&215.69&216.59&214.50\cite{KAS:barrow:1973:JCSFT2,KAS:wei:1974:JoMS}\\
    At$_2$ &2.9873&2.9799&\text{--}&117.68&118.96&\text{--}\\
    \hline\hline
  \end{tabular}
  % }
  \caption{Performance of 1eX2C and SAP-X2C Kohn-Sham with dyall-ae3z/PBE0 for coinage and halogen dimers compared to experimental values.}
  \label{tab:dft-bonddist-and-freqs}
\end{table}

\subsection{Size-Intensivity of SAP-X2C Picture Change}\label{sec:size-intensivity-test}

The 1eX2C Hamiltonian (\cref{eq:1ex2c}) does not have a thermodynamic limit due to the fact that the picture-change matrix $\mathcal{U}$ is obtained by diagonalization of the {\em core} Dirac Hamiltonian (\cref{eq:h-1b-dirac}). The latter contains the electrostatic potential of the nuclei alone, which diverges in the thermodynamic limit unless compensated by the electronic counterpart. This means that the 1eX2C picture-change matrix, and hence the 1eX2C Hamiltonian itself, do not have a thermodynamic limit. For periodic systems\cite{VRG:abraham:2024:JCTC} this can be mitigated by an ad hoc replacement of the nuclear electrostatic potential in the core Dirac Hamiltonian by the nuclear Ewald potential. Such replacement is problematic on its own since the nuclear density is not charge neutral, thus the potential itself is not uniquely defined (although the asymptotic convergence with the supercell size can be fast; see Ref. \citenum{VRG:makov:1995:PRB}). Furthermore, for non-periodic systems the fundamental issue remains.

By replacing the bare nuclear potential with SAP, the resulting SAP-Dirac Hamiltonian (\cref{eq:h-1b-sap-dirac}), the SAP-based picture-change matrices, and the SAP-X2C Hamiltonian itself have well defined thermodynamic limits. This feature critically depends on the sufficiently-rapid decay of the model atomic potentials in SAP (see \cref{eq:v-sap-asymptotics} and the associated text). In this section, we demonstrate this fact explicitly through numerical verification using fragments of a Xe crystal of increasing size.
The recipe for constructing these Xe lattice fragments and their geometries in \code{.xyz} format is provided in the Supporting Information.

To avoid the need to construct X2C Hamiltonians in the full basis, we constructed 1eX2C and SAP-X2C Hamiltonians for the single ``central'' atom $A$ embedded in a crystal fragment. Namely, we computed the \{core, SAP\} Dirac Hamiltonian in the basis of atomic spinors centered on the ``central'' atom $A$:
\begin{equation}\label{eq:h-1b-dirac-lattice}
\mathcal{H}_{A} =
    \begin{pmatrix}
    \mathbf{V}_A + \mathbf{V}^{\text{env}}_A & \mathbf{T} \\
    \mathbf{T} & \frac{\mathbf{W}_A + \mathbf{W}^{\text{env}}_A}{4c^2} - \mathbf{T}
    \end{pmatrix},
\end{equation} 
where $\mathbf{V}_A$ and $\mathbf{V}^{\text{env}}_A$ are contributions to the potential from atom $A$ and the rest of the atoms, respectively, represented in the basis of L atomic spinors on atom A:
\begin{equation}\label{eq:v-rest-bare-nuc-lattice}
    (\mathbf{V}^{\text{env}}_A)_{\mu\nu} = (  \phi_{\mu} | \sum_{B \neq A} V(r_{B})| \phi_{\nu} ).
\end{equation}
Similarly, $\mathbf{W}_A$ and $\mathbf{W}^{\text{env}}_A$ are the corresponding matrices in the basis of S atomic spinors on atom A.
The embedded Dirac Hamiltonian for atom $A$ was then used for the X2C transformation and the subsequent HF computation on the single atom $A$ as usual. Note that due to the much faster decay of contributions to $\mathbf{W}_A^\text{env}$ than to $\mathbf{V}_A^\text{env}$ ($r^{-3}$ vs. $r^{-1}$ for the core Hamiltonian case), we neglected $\mathbf{W}^{\text{env}}_A$ in these computations.

\begin{figure}
    \centering
    \includegraphics[width=0.6\textwidth]{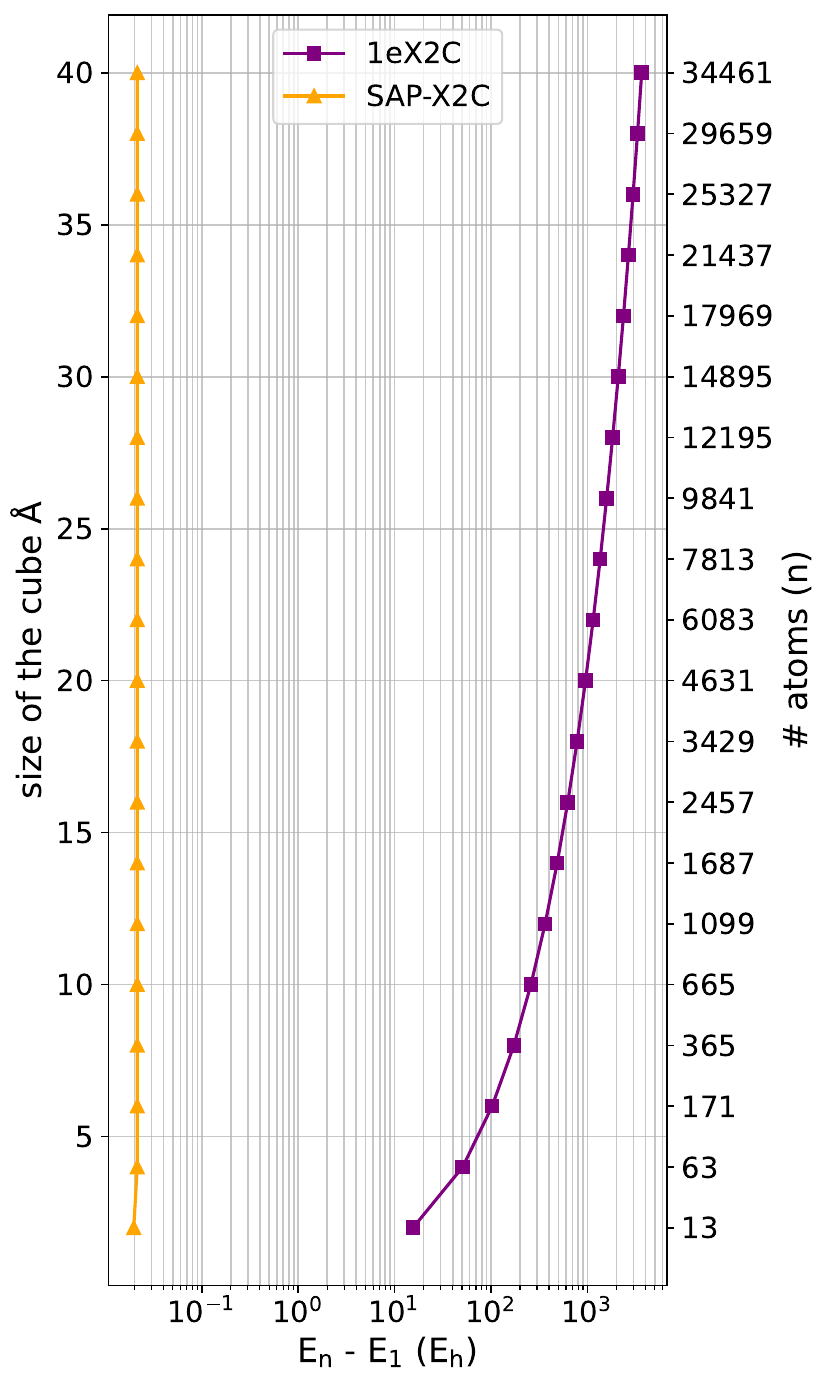}
    \caption{Difference between the X2C energies of the central Xe atom embedded in a $n$-atom crystal fragment ($E_{n}$) and an isolated Xe atom ($E_{1}$). Note the logarithmic scale of the horizontal axis.}
    \label{fig:size-intensivity-sap-vs-1ex2c}
\end{figure}

Clearly, only $\mathbf{V}^{\text{env}}_A$ and $\mathbf{W}^{\text{env}}_A$ depend on the system size; the rest of the Hamiltonian components are {\em size-intensive}.
For the {\em core} Dirac Hamiltonian of an atom at the center of a sphere of radius $L$ containing $N=\mathcal{O}(L^3)$ atoms, these contributions diverge as $\mathcal{O}(L^2)=\mathcal{O}(N^{2/3})$ and $\mathcal{O}(\log L)=\mathcal{O}(\log N)$, respectively.
This means that the 1eX2C Hamiltonian does not have a thermodynamic limit.
For the SAP Dirac Hamiltonian, both of these contributions are finite, and the corresponding SAP-X2C Hamiltonian is well behaved.
These observations are confirmed numerically; \cref{fig:size-intensivity-sap-vs-1ex2c} exhibits the energy difference between the total X2C-HF energies of the embedded and isolated atom.
The divergence of the 1eX2C energy and the perfect size-intensivity of the SAP-X2C energy are both plainly seen. 
SAP-X2C is therefore readily amenable to studies of extended and bulk systems.
This feature makes it the preferred alternative not only to 1eX2C but also to some of the more elaborate X2C flavors, such as amfX2C, which does not fully cancel out long range Coulomb interactions and requires the introduction of the more elaborate eamfX2C variant.\cite{KAS:knecht:2022:JCP}

\section{Summary}\label{sec:summary}

We introduced a simple relativistic exact 2-component (X2C) Hamiltonian that accounts for two-electron picture-change effects using their free-atom model based on Lehtola's Superposition of Atomic Potentials (SAP).\cite{KAS:lehtola:2019:JCTC,KAS:lehtola:2020:JCPa} The SAP-X2C approach preserves the low-cost and technical simplicity of the popular 1-electron X2C (1eX2C) predecessor but is significantly more accurate and has a well-defined thermodynamic limit. Specifically:
\begin{itemize}
\item The SAP-X2C is a nearly-trivial black-box extension of 1eX2C as long as derivative 3-center two-electron Coulomb integrals are available. Unlike the Atomic Mean-Field (AMF) X2C methods, SAP-X2C does not involve atomic 4C computations and avoids the need for the machinery to deal with spherical averaging, non-aufbau configurations, etc.
\item SAP-X2C is significantly more accurate than 1eX2C and should always be preferred. While SAP-X2C Hartree-Fock energies do not approximate the full 4C Dirac-Coulomb counterparts as well as the far more complex AMF X2C approaches, SAP-X2C is competitive with the AMF X2C approaches for equilibrium geometries and harmonic vibrational frequencies.
\item Unlike 1eX2C and some AMF variants, the SAP-X2C Hamiltonian has a well-defined thermodynamic limit (i.e., its matrix elements are size-intensive) and is therefore applicable to extended systems, such as periodic crystals.
\end{itemize}

Given the promising performance, far more extensive testing of SAP-X2C in its current form is clearly warranted. However, it may be possible to further improve the accuracy of SAP-X2C by ``tuning'' atomic model potentials for this purpose. This and other improvements will be explored in future work.

\begin{suppinfo}

Machine-readable XYZ files with Cartesian geometries for all species; description of the method used to construct crystal fragments used for size-intensivity tests in \cref{sec:size-intensivity-test}.

\end{suppinfo}

\begin{acknowledgement}
The authors acknowledge many insightful discussions with Trond Saue.
This work was supported by the U.S. Department of Energy via award DE-SC0022327. The development of the \code{Libint}  software library is supported by the Office of Advanced Cyberinfrastructure, National Science Foundation (Award OAC-2103738).
The authors acknowledge Advanced Research Computing at Virginia Tech \\({\tt https://arc.vt.edu/}) for providing computational resources and technical support that have contributed to the results reported within this paper.
\end{acknowledgement}

\appendix
\appendixpage

\section{Notation}\label{sec:notation}

The Fock-space Hamiltonian $H \equiv \, h + g$
consists of 1-particle ($h$) and 2-particle ($g$) components defined as:
\begin{align}
    h \equiv & \, h^p_q a_q^\dagger a_p ;\\
    g \equiv & \, \frac{1}{2} g^{pq}_{rs} a_r^\dagger a_s^\dagger a_q a_p
\end{align}
where fermionic creator/annihilator $a^\dagger/a_p$ creates a particle in a single-particle (sp) state $\ket{\phi_p}$; indices $p,q,r$ and $s$ represent the sp states, and in this work, they are either spinors (2C) or bispinors (4C).
$h^p_q$ and $g^{pq}_{rs}$
are the matrix elements of the 1- and 2-body parts of the Hamiltonian:
\begin{align}
    h^p_q \equiv & \, \int \phi_q^*(1) \hat{h}(1) \, \phi_q(1) \mathrm{d}1 \\
    g^{pq}_{rs} \equiv & \, \int \phi_r^*(1) \phi_s^*(2) \hat{g}(1,2) \phi_p(1) \phi_q(1) \, \mathrm{d}1 \, \mathrm{d}2.
\end{align}

It is convenient to rewrite $H$ in a normal-ordered form with respect to the reference Slater determinant state $\ket{\Phi}$:
\begin{align}
    H = & E_0 + f + w
\end{align}
where $E_0 \equiv \bra{\Phi}H\ket{\Phi}$ is the reference energy, and $f$ and $w$ are the 1-body (Fock operator) and 2-body components of the normal-ordered Hamiltonian $\{H\}$, defined as:
\begin{align}
    f \equiv & f^p_q \left\{ a_q^\dagger a_p \right\} \\
    w \equiv & \frac{1}{2} g^{pq}_{rs} \left\{ a_r^\dagger a_s^\dagger a_q a_p \right\}
\end{align}
where $\{\cdot\}$ denotes normal-ordering and
\begin{align}
f^p_q \equiv & h^p_q + \left(g^{pi}_{qi} - g^{pi}_{iq}\right)
\end{align}
are the elements of the Fock matrix and the index $i$ represents the occupied sp states.

Matrix representations of 1-body operators in (electronic) spinor and bispinor basis will be represented in boldface and calligraphic fonts, respectively, e.g.:
\begin{align}
    h_p^q = \begin{cases}
    (\mathcal{H})_p^q, & \text{4C}, \\
    (\mathbf{H})_p^q, & \text{2C}.
     \end{cases}
\end{align}
The projection from the bispinor to the upper spinor basis is accomplished by
\begin{align}
P_\text{e} \equiv \,
\begin{pmatrix}
1 & 0 & 0 & 0 \\
0 & 1 & 0 & 0 
\end{pmatrix}.
\end{align}
Thus, $\mathbf{H} = P_\text{e} \mathcal{H} P_\text{e}^\dagger$.

\bibliography{vrgrefs, KAS, misc}

\begin{tocentry}

\includegraphics[width=3.25in]{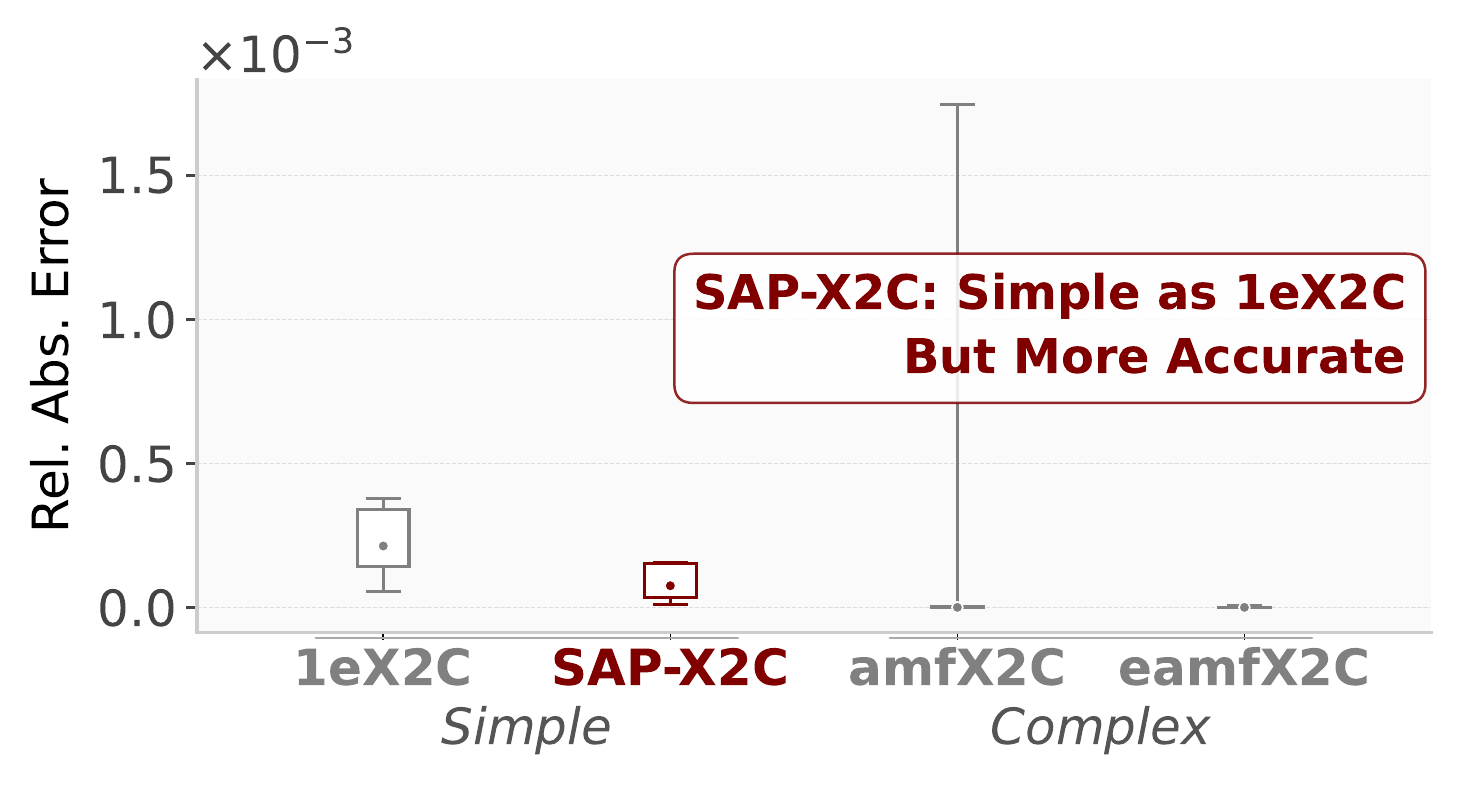} 

\end{tocentry}

\end{document}

\date{\today}
\maketitle

\section{Molecular geometries}\label{sec:mol-geoms}
The Supporting Information accompanying this publication includes a .zip file that
contains:
\begin{enumerate}
\item Files containing all molecular geometries used for the energy calculations in this work, in \code{.xyz} format.
\item \code{.xyz} files for Xe lattice fragments. The recipe to construct these fragments is described below in \cref{sec:Xe-frag-construct}.
\end{enumerate}

\section{Construction of Xe lattice fragments}\label{sec:Xe-frag-construct}
Xe forms a face-centered cubic (FCC) lattice structure with a unit cell of length $a = 6.2023$ \AA \cite{KAS:sears:1962:JCP}. 
The construction of crystal fragments follows a simple recipe, and can primarily be formed by an alternating arrangement of $yz$ plane tiling from \cref{fig:xe-fragment-layer1,fig:xe-fragment-layer2} along x-axis. Let us first look at \cref{fig:xe-fragment-layer1} (L1) that tiles $yz$ planes at a given $x$. The distance between any two Xe atoms along the $y$ or $z$ axis is the unit cell size $a$, and the adjacent (parallel) 1D arrangements (or 1D layers) are separated by $a/2$, depicted by the dotted lines in \cref{fig:xe-fragment-layers}. In \cref{fig:xe-fragment-layer1}, it can be seen that the smallest square formed by the dotted lines around the origin constitutes the face of the FCC unit cell. \cref{fig:xe-fragment-layer2} shows the second type of arrangement (L2) that tiles $yz$ planes needed for the FCC lattice construction. Now, if we arrange L1 at $x=0$, L2 at $x=a/2$ and L1 again at $x=a$, it can be seen that the cube $(0,0,0)\rightarrow (a,a,a)$ forms the FCC unit cell. In this work, to keep the arrangement of atoms symmetric around the atom at the origin, we construct the fragments by arranging L1s at $x = Na$ and L2s at $x=(N+1)a/2$, where $n\in .... -2,-1, 0, 1,2,...$ , and then taking cubic slices between $(-Na/2,-Na/2,-Na/2)\rightarrow (Na/2,Na/2,Na/2)$. 
\begin{figure}
\centering
\begin{subfigure}[b]{0.6\textwidth}
    \centering
    \begin{tikzpicture}[scale=1.0]
    
        % parameters
        \def\N{2}          % number of atoms per row/column
        \def\step{1.5}     % lattice spacing
        \def\R{0.3}        % atom radius
    
        % bold y- and z-axes through origin
        \draw[very thick,->] (-\N*\step - \step,0) -- (\N*\step + \step,0) node[right] {$y$};
        \draw[very thick,->] (0,-\N*\step - \step) -- (0,\N*\step + \step) node[above] {$z$};    
        
        % dotted grid lines
        \draw[step=\step, dotted, gray, line width=1pt] 
            (-\step*\N - 1,-\step*\N -1) grid (\step*\N+1,\step*\N+1);
        
        \filldraw[fill=blue!20,draw=blue]  (-3.0, -3.0) circle (\R) node {\small Xe};
        \filldraw[fill=blue!20,draw=blue]  (-3.0, 0.0) circle (\R) node {\small Xe};
        \filldraw[fill=blue!20,draw=blue]  (-3.0, 3.0) circle (\R) node {\small Xe};
        \filldraw[fill=blue!20,draw=blue]  (-1.5, -1.5) circle (\R) node {\small Xe};
        \filldraw[fill=blue!20,draw=blue]  (-1.5, 1.5) circle (\R) node {\small Xe};
        \filldraw[fill=blue!20,draw=blue]  (0.0, -3.0) circle (\R) node {\small Xe};
        \filldraw[fill=blue!20,draw=blue]  (0.0, 0.0) circle (\R) node {\small Xe};
        \filldraw[fill=blue!20,draw=blue]  (0.0, 3.0) circle (\R) node {\small Xe};
        \filldraw[fill=blue!20,draw=blue]  (1.5, -1.5) circle (\R) node {\small Xe};
        \filldraw[fill=blue!20,draw=blue]  (1.5, 1.5) circle (\R) node {\small Xe};
        \filldraw[fill=blue!20,draw=blue]  (3.0, -3.0) circle (\R) node {\small Xe};
        \filldraw[fill=blue!20,draw=blue]  (3.0, 0.0) circle (\R) node {\small Xe};
        \filldraw[fill=blue!20,draw=blue]  (3.0, 3.0) circle (\R) node {\small Xe};
    
    \end{tikzpicture}
    \caption{Layer type-1 of Xe atoms in the $yz$-plane (L1).}
    \label{fig:xe-fragment-layer1}
\end{subfigure}
\hfill
    \begin{subfigure}[b]{0.6\textwidth}
    \centering
    \begin{tikzpicture}[scale=1.0]
    
        % parameters
        \def\N{2}          % number of atoms per row/column
        \def\step{1.5}     % lattice spacing
        \def\R{0.3}        % atom radius
    
        % bold y- and z-axes through origin
        \draw[very thick,->] (-\N*\step - \step,0) -- (\N*\step + \step,0) node[right] {$y$};
        \draw[very thick,->] (0,-\N*\step - \step) -- (0,\N*\step + \step) node[above] {$z$};    
        
        % dotted grid lines
        \draw[step=\step, dotted, gray, line width=1pt] 
            (-\step*\N - 1,-\step*\N -1) grid (\step*\N+1,\step*\N+1);
        
        \filldraw[fill=blue!20,draw=blue]  (-3.0, -1.5) circle (\R) node {\small Xe};
        \filldraw[fill=blue!20,draw=blue]  (-3.0, 1.5) circle (\R) node {\small Xe};
        \filldraw[fill=blue!20,draw=blue]  (-1.5, -3.0) circle (\R) node {\small Xe};
        \filldraw[fill=blue!20,draw=blue]  (-1.5, 0.0) circle (\R) node {\small Xe};
        \filldraw[fill=blue!20,draw=blue]  (-1.5, 3.0) circle (\R) node {\small Xe};
        \filldraw[fill=blue!20,draw=blue]  (0.0, -1.5) circle (\R) node {\small Xe};
        \filldraw[fill=blue!20,draw=blue]  (0.0, 1.5) circle (\R) node {\small Xe};
        \filldraw[fill=blue!20,draw=blue]  (1.5, -3.0) circle (\R) node {\small Xe};
        \filldraw[fill=blue!20,draw=blue]  (1.5, 0.0) circle (\R) node {\small Xe};
        \filldraw[fill=blue!20,draw=blue]  (1.5, 3.0) circle (\R) node {\small Xe};
        \filldraw[fill=blue!20,draw=blue]  (3.0, -1.5) circle (\R) node {\small Xe};
        \filldraw[fill=blue!20,draw=blue]  (3.0, 1.5) circle (\R) node {\small Xe};
    
    \end{tikzpicture}
    \caption{Layer type-2 of Xe atoms in the $yz$-plane (L2).}
    \label{fig:xe-fragment-layer2}
\end{subfigure}
\caption{}
\label{fig:xe-fragment-layers}
\end{figure}

\bibliography{vrgrefs, KAS, misc}